\documentclass[fleqn,usenatbib]{mnras}

\usepackage{newtxtext,newtxmath}

\usepackage[T1]{fontenc}

\DeclareRobustCommand{\VAN}[3]{#2}
\let\VANthebibliography\thebibliography
\def\thebibliography{\DeclareRobustCommand{\VAN}[3]{##3}\VANthebibliography}

\usepackage{graphicx}	
\usepackage{amsmath}	

\newcommand{\tbn}{$\theta_{Bn}$ }	
\newcommand{\rd}{$^{\mathrm{rd}}$ }

\title[Shocklets at IP shock]{Multi-spacecraft observations of shocklets at an interplanetary shock.}

\author[D. Trotta et al.]{
D. Trotta,$^{1}$\thanks{E-mail: d.trotta@imperial.ac.uk }
H. Hietala,$^{1,2}$
T. Horbury,$^{1}$
N. Dresing,$^{3}$
R. Vainio,$^{3}$
L. Wilson III,$^{4}$
I. Plotnikov,$^{5}$
E. Kilpua$^{6}$
\\
$^{1}$ The Blackett Laboratory, Department of Physics, Imperial College London, London SW1234, UK \\
$^{2}$ School of Physics and Astronomy, Queen Mary University of London, London E1 4NS, UK \\
$^{3}$ Department of Physics and Astronomy, University of Turku, Turku, \\
$^{4}$ NASA Goddard Space Flight Center, Greenbelt, MD, United States \\
$^{5}$ IRAP, Université Toulouse III – Paul Sabatier, CNRS, CNES, Toulouse, France \\
$^{6}$  Department of Physics, University of Helsinki, Helsinki, Finland 
}

\date{Accepted XXX. Received YYY; in original form ZZZ}

\pubyear{2022}

\begin{document}
\label{firstpage}
\pagerange{\pageref{firstpage}--\pageref{lastpage}}
\maketitle

\begin{abstract}
    Interplanetary (IP) shocks are fundamental building blocks of the heliosphere, and the possibility to observe them \emph{in-situ} is crucial to address important aspects of energy conversion for a variety of astrophysical systems. Steepened waves known as shocklets are known to be important structures of planetary bow shocks, but they are very rarely observed related to IP shocks. We present here the first multi-spacecraft observations of shocklets observed by upstream of an unusually strong IP shock observed on November 3rd 2021 by several spacecraft at L1 and near-Earth solar wind. The same shock was detected also by radially aligned Solar Orbiter at 0.8 AU from the Sun, but no shocklets were identified from its data, introducing the possibility to study the environment in which shocklets developed. The Wind spacecraft has been used to characterise the shocklets, associated with pre-conditioning of the shock upstream by decelerating incoming plasma in the shock normal direction. Finally, using the Wind observations together with ACE and DSCOVR spacecraft at L1, as well as THEMIS B and THEMIS C in the near-Earth solar wind, the portion of interplanetary space filled with shocklets is addressed, and a lower limit for its extent is estimated to be of about 110 $R_E$ in the shock normal direction and 25 $R_E$ in the directions transverse to the shock normal. Using multiple spacecraft also reveals that for this strong IP shock, shocklets are observed for a large range of local obliquity estimates (9-64 degrees).
    
\end{abstract}




\section{Introduction}
\label{sec:introduction}

Shocks are ubiquitous, and they are fundamental for a broad range of astrophysical systems \citep[e.g.,][]{Kivelson1995,Bykov2019}. Generally speaking, shocks convert directed flow energy (upstream) into heat and magnetic energy (downstream) and, in the collisionless case, in energetic particles \citep[e.g.,][]{Burgess2015}.

Interplanetary (IP) shocks, found in the heliosphere, are generated as a consequence of solar  phenomena, such as Coronal Mass Ejections (CME) and Stream Interaction Regions (SIR) \citep[][]{Gosling1974,Dessler1963, Kilpua2017, Richardson2018}. IP shocks play an important role for the overall heliosphere energetics, due to their ability to accelerate particles to high energies and modify the plasma environments in their surroundings \citep[see][for a review]{Reames1999}. Furthermore, IP shocks provide a unique opportunity for \emph{in-situ} observations using the the instrumentation on board of spacecraft, a mean of analysis inaccessible in astrophysical shocks. Another group of shocks routinely observed in the heliosphere are the planetary bow shocks, resulting from the interaction between the supersonic solar wind and the planets that behave as obstacles \citep[][]{Hoppe1982}. From this point of view, the Earth's bow shock has become a prototype for studying various phenomena characterised by the presence of shocks, due to the convenience to be probed, starting from the early \emph{Pioneer} evidences \citep[][]{Dungey1979} to the modern spacecraft observations such as the Magnetospheric MultiScale (MMS) mission \citep{Burch2016}.Generally speaking, IP shocks are weaker and show larger radii of curvature with respect to planetary bow shocks, a feature inducing several differences is their upstream/downstream plasma environments \citep[e.g.,][]{Kilpua2015,Dresing2016,Eastwood2015, Wilson2016ch}. 

\begin{table*}
	\centering
	\caption{Shock arrival time and parameters computed for different spacecraft. The parameters shown are (left to right): shock normal vector, \tbn, magnetic compression ratio $\rm{r}_B$, gas compression ratio $\rm{r}$, shock speed $v_{\mathrm{sh}}$, upstream plasma beta $\beta_{\mathrm{up}}$, fast magnetosonic and Alfv\'enic Mach numbers ($\rm{M_{\rm{fms}}}$ and $\rm{M_{\rm{A}}}$, respectively).   The shock normals are shown in the GSE frame of reference, with \tbn expressed in degrees. The shock speed $v_{\mathrm{sh}}$ in the spacecraft frame is expressed in km/s and it is aligned to the shock normal.}
	\label{tab:tab_event}
	\begin{tabular}{lcccccccccr} 
		\hline
		Spacecraft & GSE position [$R_{E}$]&  Shock Time [UT] & $\langle \hat{\mathrm{n}}_{\mathrm{GSE}} \rangle$ &$\langle \theta_{Bn}\rangle$ [$^\circ$] & $\langle \rm{r}_B \rangle$ & $\langle \rm{r} \rangle$  & $\langle v_{\rm{sh}} \rangle$ &  $\beta_{\mathrm{up}}$& $\rm{M_{\rm{fms}}}$ & $\rm{M_{\rm{A}}}$ \\
		\hline
		Solar Orbiter & [3482.9, 283.9. -744.8]& 14:04:26 & [-0.51,  0.49, -0.71] & 45.3 & 2.62 & 1.47 & 691.8 & 0.5 &5.5 & 6.2 \\
        Wind & [196.5, 14.3, -10.5]& 19:35:01 &  [-0.87, -0.04, -0.49] &33.1 & 3.10 & 5.15 & 768.8 & 0.4 &5.3 & 5.6 \\
        ACE  & [230.9, -40.0, 12.49]& 19:24:05 &   [-0.67,   0.33, -0.66] & 9.6 & 1.93 & - & - & - & - & - \\
        DSCOVR & [240.2, 30.9, 25.5] &19:24:50 &  [-0.67,  0.03, -0.74] & 13.3 & 2.83 & - & - & - & - & - \\
        THB  & [52.9, -12.8, 2.8] &19:43:20 & [-0.79, -0.01, -0.61] & 50.1 & 2.06 & 2.93 & 625.9 & - & - & 5.8 \\
        THC  & [54.7, -15.7, 2.9] &19:43:30 & [-0.79,  0.01, -0.61] & 64 & 1.2 & 1.54 & 686.1 & - & - & 6.4 \\

		\hline
	\end{tabular}
\end{table*}

The shock structure and behaviour is regulated by several parameters, one of the most important of which is the angle between the shock normal direction and the upstream magnetic field, \tbn. When \tbn is close to 90$^\circ$, the shock is quasi-perpendicular. On the other hand, for \tbn values close to 0$^\circ$ (corresponding to an upstream magnetic field almost normal to the shock surface), the shock is quasi-parallel.  Other important parameters for the shock behaviour are the shock Alfv\'enic and fast magnetosonic Mach numbers, i.e., the ratio between the shock speed in the upstream flow frame and the upstream Alfv\'en and fast magnetosonic speed, respectively ($M_{\rm{A}} \equiv v_{\rm{sh}}/v_{\rm{A}}$ and $M_{\rm{fms}} \equiv v_{\rm{sh}}/v_{\rm{fms}}$). Finally, another important parameter to address shock behaviour is the plasma beta, expressed as a ratio between thermal and Alfv\'en speeds $\beta \equiv v_{\rm{th}}^2/ v_{\rm{A}}^2$. Particle reflection and subsequent propagation far upstream is favoured at high Mach number (supercritical) quasi-parallel shocks \citep[][]{Kennel1985}. This  introduces the possibility for reflected particles to interact with the upstream plasma over long distances, creating unstable distributions and a collection of disturbances in the plasma properties. This region of interaction between the shock and its upstream is called the foreshock, and its fundamental for many aspects of energy conversion in collisionless plasmas \citep[][]{Wilson2016}.



Shock reflected, energetic ions are also thought to be responsible for the emergence of steepened waves in the shock upstream. These steep structures, that are observed to shave short (< 1 min) duration, are called shocklets and Short Large Amplitude Magnetic Structures (SLAMS) depending on their typical signatures, and are both characterised by steep, strong enhancements of the magnetic field magnitude \citep[e.g.,][]{Stasiewicz2003, Plaschke2018}. Shocklets are likely to play an important role for particle acceleration at quasi-parallel shocks, due to their ability to induce effective pre-conditioning of incoming plasma before its interaction with the shock \citep[e.g.,][]{Wilson2013}. Despite many observational \citep[e.g.,][]{Lucek2008} and theoretical \citep[e.g.,][]{Hellinger1996,Scholer2003} efforts, the (nonlinear) mechanisms leading to the formation of shocklets are still a matter of debate. The emerging picture is that shocklets are created in consequence of Ultra-Low-Frequency (ULF) foreshock waves  \citep[see][]{Lucek2002,Lucek2008, Wilson2016ch}.

\begin{figure}
	\includegraphics[width=\columnwidth]{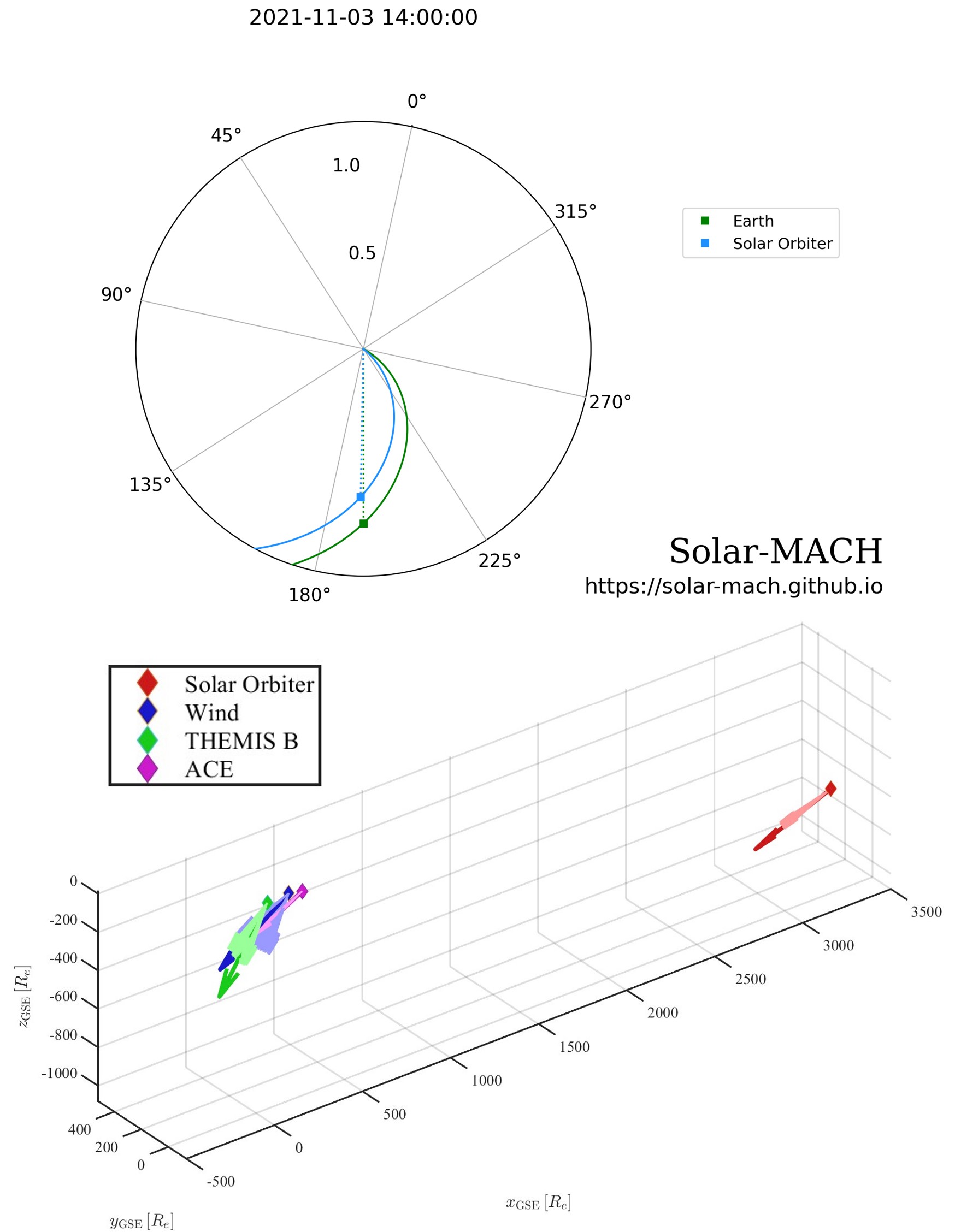}
    \caption{\emph{Top}: Spacecraft configuration at 14:00 on November 3$^{rd}$ 2021. The Sun is at the centre of the plot, with radial and magnetic field connections to Solar Orbiter and Earth represented by dotted and solid lines, respectively. \emph{Bottom}: Three-dimensional overview of the Solar Orbiter, Wind, THEMIS B and ACE spacecraft relative positions (points). Earth is located in (0,0,0). Superimposed are the shock normals computed using different averaging windows for each spacecraft, and their mean (light and dark arrows, respectively). }
    \label{fig:fig1_Orbit}
\end{figure}

Despite the mystery surrounding their formation mechanisms, shocklets have been observed \emph{in-situ} for a large variety of shocks. These include the planetary bow shocks of Earth \citep[e.g.,][]{Russell1971}, Jupiter \citep{Tsurutani1993} and Saturn \citep[][]{Bertucci2007,Andres2013}, as well as the cometary bow shocks of the Giacobini-Zinner \citep{Tsurutani1987,Thomsen1986}, Halley \citep{Naeem2020} and Grigg-Skjllerup \citep{Coates1997}. 

Most of the knowledge about shocklets is due to  observations at planetary bow shocks, where they are defined as having weak magnetic compression $\delta B/B_0 \lesssim 2$ and a duration of around 30 s at Earth's bow shock. Shocklets are characterised by an upstream sharp leading edge followed by a slower relaxation \citep[e.g.,][]{Lucek2002}. Futhermore, it has been found that shocklets are often associated with whistler wave precursors \citep{Hoppe1981} unstable electron distributions \citep{Wilson2009}. Importantly, all shocklets observations show diffuse ion distributions \citep[e.g.,][]{Hoppe1983}, which likely represent an important ingredient for efficient wave steepening, as shown by early simulation works \citep[e.g.][]{Omidi1990}. 

\begin{figure*}
	\includegraphics[width=\textwidth]{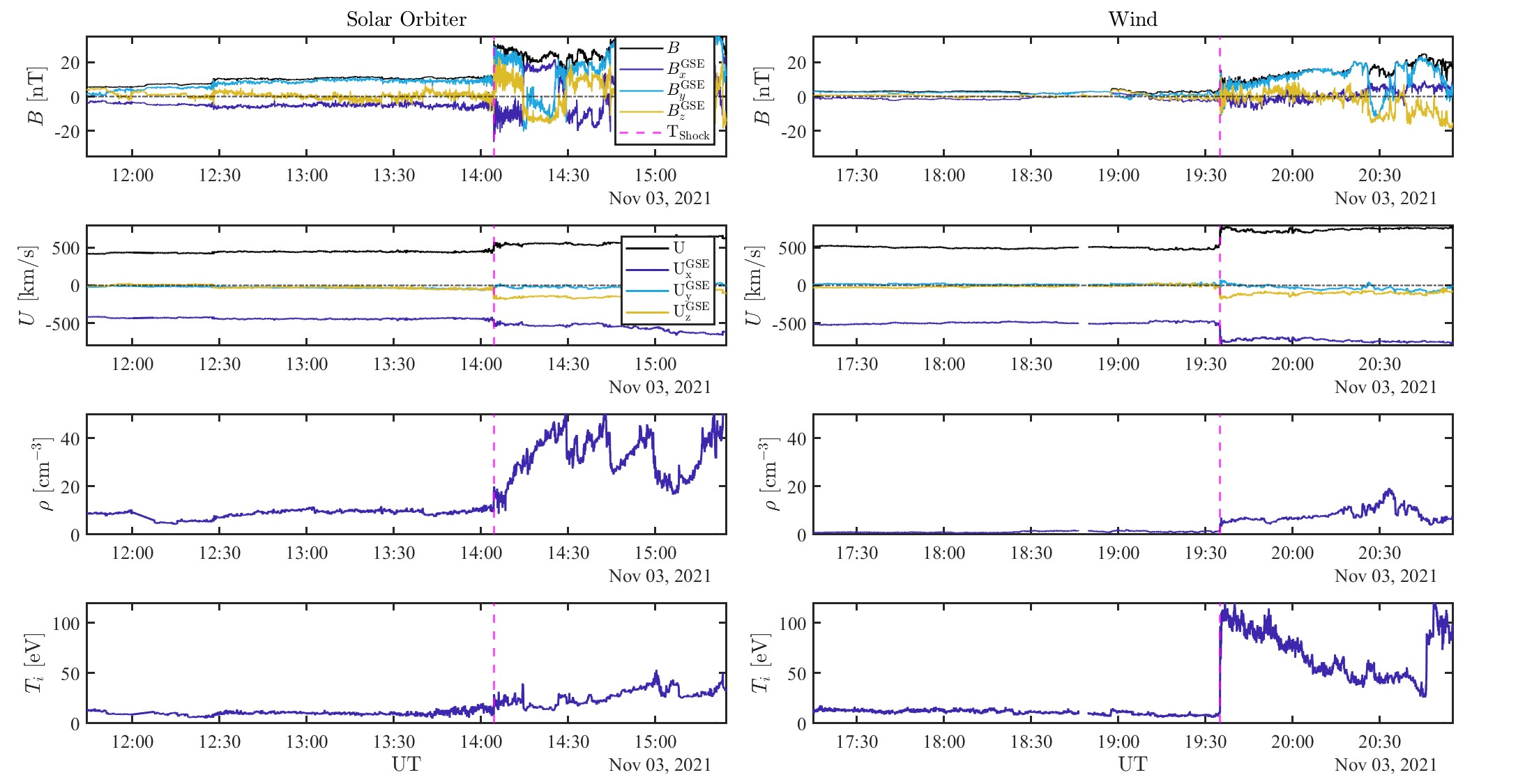}
	\caption{Magnetic field magnitude and components, ground-computed ion bulk flow speed, ion density, ion temperature (top to bottom) for the IP shock observed first by Solar Orbiter (left) and later by Wind (right).}
	\label{fig:fig2_SoloVsWind}
\end{figure*}

At IP shocks, shocklets observations are much more rare than planetary bow shocks, making the few available observations of particular interest, as they yield to a better understanding of the origin and evolution of these phenomena. The first of these observations is due to \citet{Lucek1997}, who reported the presence of a  structure in the magnetic field measured by the \emph{Ulysses} spacecraft \citep{Balogh1995} upstream of a quasi-parallel interplanetary shock on 06 January 1992, with features similar to shocklets observed at Earth's bow shock by \citet{Le1992}, with the difference of being associated with whistler precursor. Unfortunately, this observation was limited by the fact that only magnetic field data was available, leaving the plasma properties around the structure out of reach. More than a decade later, using the Wind spacecraft data, \citet{Wilson2009} reported the presence of 12 shocklets upstream of the quasi-perpencdicular, high Mach number IP shock of 6 April 2000. This event, labelled by the authors ``the unusual event'' (with respect to more than 400 other IP shocks observed by Wind), is the second (and last before the present work, to the best of our knowledge) shocklet observation at IP shocks. To make the 6 April 2000 event even more interesting is the fact that the local \tbn of the shock was estimated to be of 68$^\circ$, and so less likely to have upstream conditions favourable for wave steepening.

In this work, we present the first multi-spacecraft observations of IP shock shocklets (seen by several spacecraft near Earth), enabling us to investigate the spatial and temporal transient nature of such steep waves. Furthermore, the same shock is observed by Solar Orbiter, well aligned radially and  $\sim$ 3500 Earth radii upstream of Earth. At Solar Orbiter, no shocklets are found, making it possible to study the environment in which upstream wave steepening happened.

The paper is organised as follows: Section \ref{sec:data_methods} presents the spacecraft data products employed in this work (Sec. \ref{subsec:data_in_situ}) and the techniques used for shock parameter estimation and shocklets characterisation (Sec. \ref{subsec:methods_parameters}); the spacecraft observations are presented in Section~\ref{sec:sc_Observ}, with an overview of the event shown in Section~\ref{subsec:overview}; the detailed shocklet observations are presented in Section~\ref{subsec:shocklets}, and the multi-spacecraft observations of such structures are then presented in Section~\ref{subsec:multi-sc}; the paper ends with the conclusions reported in Section~\ref{sec:discussion}.

\section{Data and methods}
\label{sec:data_methods}
\subsection{\emph{In-situ} measurements}
\label{subsec:data_in_situ}
Throughout this study, magnetic field and plasma data from several spacecraft have been used. At Solar Orbiter, the magnetic field has been measured with a resolution of 64 vectors$/$s by the flux-gate magnetometer MAG \citep{Horbury2020}, while ion bulk flow, density and temperature are the ground computed plasma moments measured by the Solar Wind Analyser (SWA) suite \citep{Owen2020}, with a 3 seconds resolution. 

 For the Wind data shown \citep{Wilson2021}, the magnetic field is measured using the Wind Magnetic Field Investigation (MFI), at a resolution of 11 vectors$/$s \citep{Lepping1995}, and the ion moments are obtained using the Wind Three-Dimensional Plasma and Energetic Particle Investigation (3DP) instrument at 3 seconds resolution \citep{Lin1995}. The measurements obtained with the 3DP instrument are consistent with those obtained with the Solar Wind Experiment instrument \citep{Ogilvie1995}. A collection of other spacecraft has been used near Earth, in particular: the Advanced Composition Explorer (ACE) magnetic field experiment  at 1 s resolution \citep{Smith1998}, the Deep Space Climate Observatory (DSCOVR) magnetometer at 1 second resolution \citep{Szabo2016}, and the THEMIS B and C Flux Gate Magnetometer \citep[FGM,][]{Auster2008} and electrostatic analyser \citep[ESA][]{Mcfadden2008} at 4 seconds resolution. The high resolution provided by Wind data has been used to investigate the fine details of shocklets (including their high-frequency wave precursors), as discussed below.

\subsection{Shock parameter estimation and shocklet characterisation}
\label{subsec:methods_parameters}

The shock normal (and therefore the \tbn) estimation is done using the mixed mode 3 method \citep[MX3, see][]{Paschmann2000} for Solar Orbiter, Wind and THEMIS B and C. The results obtained with the other mixed modes are compatible with the ones shown here. When plasma data for the event is not available (as in the case of DSCOVR and ACE), the magnetic coplanarity method is used to determine the shock normal vector. The shock speed is computed through the mass flux conservation, and it is along the shock normal, in the spacecraft frame. Such techniques for shock parameter estimation have been extensively used and discussed in previous literature \citep[e.g.][]{Koval2008}. Given the nature of such techniques, care has been taken to choose appropriate time intervals to define upstream and downstream of the shock at each observation. For all the parameters computed and presented here, the averaging $\langle \rangle$ indicates that different upstream/downstream windows have been used systematically, to make sure that the parameters estimation is robust. Here, we used a range of upstream/downstream averaging windows lasting in a range between $\sim$30 secs to 5 minutes. The properties and importance of this systematic way of computing shock parameters starting from a single spacecraft crossing signal will be an object of a separate study.

A summary of the parameters estimated for each spacecraft crossing can be found in Table~\ref{tab:tab_event}. As it will be discussed in detail below, the shock appears to have an unusually high Mach number, a key property to address its behaviour. Furthermore, with reference to Table~\ref{tab:tab_event}, we note that the parameter estimation at THB and THC is the most sensitive to the choice of upstream and downstream windows, due to the strong structuring of the shock transition and the resolution available for the measurements. The parameter estimation involving temperature measurements for the THEMIS B and C spacecraft have been discarded here, due to the fact that such measurements are known to overestimate ion temperature in the solar wind, when the spacecraft are in magnetospheric mode.

 Another important part of the methods used in this work has to do with shocklet identification. Here, shocklets are identified by visual inspection as steep enhancements of magnetic field with $\delta B/B_0 \lesssim 2$, with an upstream sharp leading edge followed by a slower relaxation, following the definition in \citet{Wilson2016ch}. In absence of a more formal definition of shocklet structure accepted by the literature, throughout this work we identify them by visual inspection, looking for the features discussed above (see Section~\ref{sec:introduction}).

\section{Spacecraft Observations}
\label{sec:sc_Observ}

\subsection{Event Overview}
\label{subsec:overview}

On November 3\rd 2021, a fast-forward, CME-driven IP shock reached Solar Orbiter. Later on, the shock was observed by several other spacecraft near the Earth. Table \ref{tab:tab_event} shows the shock arrival time and key parameters as observed by the Solar Orbiter, Wind, DSCOVR, ACE, THEMIS B and THEMIS C (columns). 

Given the large number of satellites observing this shock, we checked if the high shock speeds computed using local upstream and downstream averaging windows and the mass flux algorithm are compatible with multiple spacecraft timing techniques. To this end, the shock speed was estimated using the local shock normals and using a two-spacecraft timing, as well as with a four-spacecraft timing technique using the near-Earth spacecraft \citep[see][for details]{Paschmann2000}. These estimations are compatible with what we found using the mass flux algorithm, but yield to a large spread of values with $450 \lesssim v_{\mathrm{sh}} \lesssim 1100 \, \mathrm{km/s}$, suggesting that the assumption of shock planarity may not be well suited for this event. Other effects responsible for this large spread of values may be related to rotation, curvature and mass-loading effects.

The top panel of Figure~\ref{fig:fig1_Orbit} shows an overview of the Sun - Earth system, including the position of Solar Orbiter. Here, the dashed lines show the radial connection between Solar Orbiter (the Earth) and the Sun, while the solid lines represent the connections through the Parker spiral. It can be noted that Solar Orbiter is well radially-aligned to Earth (and L1), making it a relevant configuration for multi-spacecraft studies. 

This interesting alignment has been put in the context on the IP shock passage in the bottom panel of Figure~\ref{fig:fig1_Orbit}. Here, the positions of Solar Orbiter, Wind, THEMIS B and ACE are shown in the three-dimensional space in $\mathrm{GSE}$ coordinates (diamonds). Superimposed to the spacecraft position are the shock normal vectors, computed as described in Section~\ref{subsec:methods_parameters}. The dark arrows represent the average shock normal vectors. It is possible to note that the normals computed for the spacecraft at L1 have some degree of fluctuation, probably due to shock front irregularities, a typical feature of high Mach number, supercritical shocks \citep[e.g.,][]{Kajdic2019}, consistent with the parameter estimation in Table~\ref{tab:tab_event}.

\begin{figure}
	\includegraphics[width=\columnwidth]{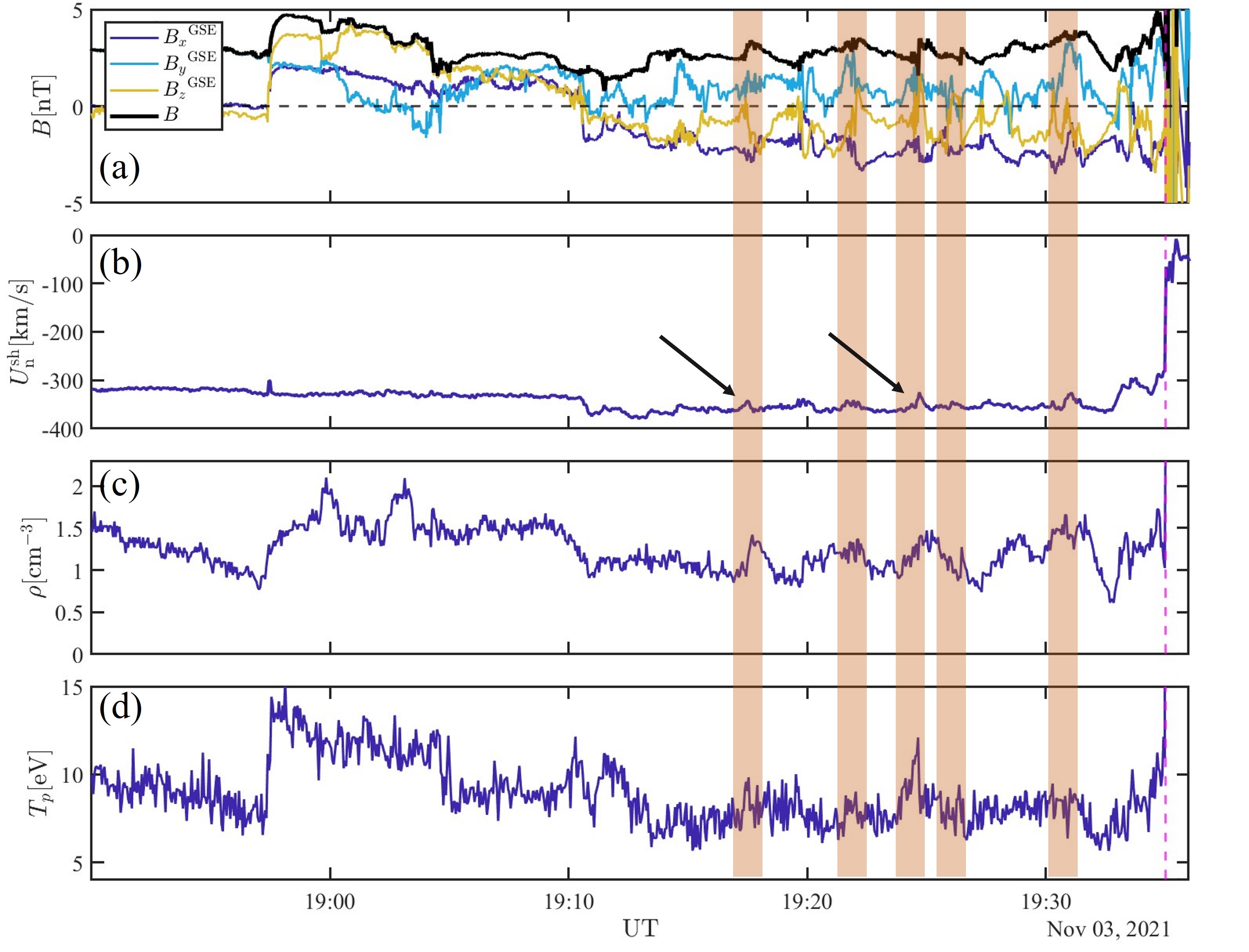}
    \caption{The shock upstream observed by the Wind spacecraft. From top to bottom: magnetic field magnitude (black) and its components in the GSE coordinate system (colors), ion bulk flow speed normal to the shock, in the shock rest frame, proton density and temperature observed by the Wind-3DP instrument. The arrows mark the upstream proton bulk flow speed deceleration correspondent to two of the most clear shocklets, discussed in detail in Figure~\ref{fig:fig4_hodo}. The dashed magenta line marks the shock arrival.}
    \label{fig:fig3_shocklets_oview}
\end{figure}

Figure~\ref{fig:fig2_SoloVsWind} shows the \emph{in-situ} observations of the November 3\rd IP shock as seen by Solar Orbiter (left) and Wind (right). As shown in Table~\ref{tab:tab_event}, the observed shock parameters indicate that we are in presence of a shock with a \tbn that has strong local variations ($9 \lesssim \theta_{Bn} \lesssim 64$). The shock has unusually high Mach numbers ($\mathrm{M}_{\mathrm{fms}} \sim 5$, $\mathrm{M}_{\mathrm{A}} \sim 6$) with respect to other IP shocks~\citep[see][for example]{Kilpua2015}. 

The IP shock crossed the Wind spacecraft around 19:35. For this crossing, the local estimation of the shock normal vector using magnetic coplanarity and mixed modes are all consistent, and indicate a low \tbn value (of about 33$^\circ$). These Wind observations are shown on the right hand side panels of Figure~\ref{fig:fig2_SoloVsWind}. The magnetic field observed by Wind shows very extended structuring over a broad range of scales both upstream and downstream, consistent with the quasi-parallel geometry inferred for the shock \citep[e.g.,][]{BlancoCano2016}. The fast magnetosonic and Alfv\'enic Mach numbers are high also at Wind, (5.3 and 5.6, respectively). For this event, we observe a very small value for the upstream proton density (about 2 particles per cm$^{-3}$, a value consistent for both the 3DP and SWE experiment on-board the spacecraft), with the gas compression ratio exceeding the MHD limiting value of 4 (with $\langle \mathrm{r} \rangle \sim 5.15$). An extended range of fluctuations is also observed for the plasma moments, as expected for this strong shock. The ion density and temperature increase sharply upon the shock arrival, and are modulated by the large scale downstream fluctuations. At Wind, temperature is much higher than at Solar Orbiter, and vice-versa for the density. This feature is interesting, especially as the spacecraft are well-aligned radially. To investigate the reasons for such a behaviour is beyond the scope here and will be object of further investigation.

\begin{figure*}
	\includegraphics[width=\textwidth]{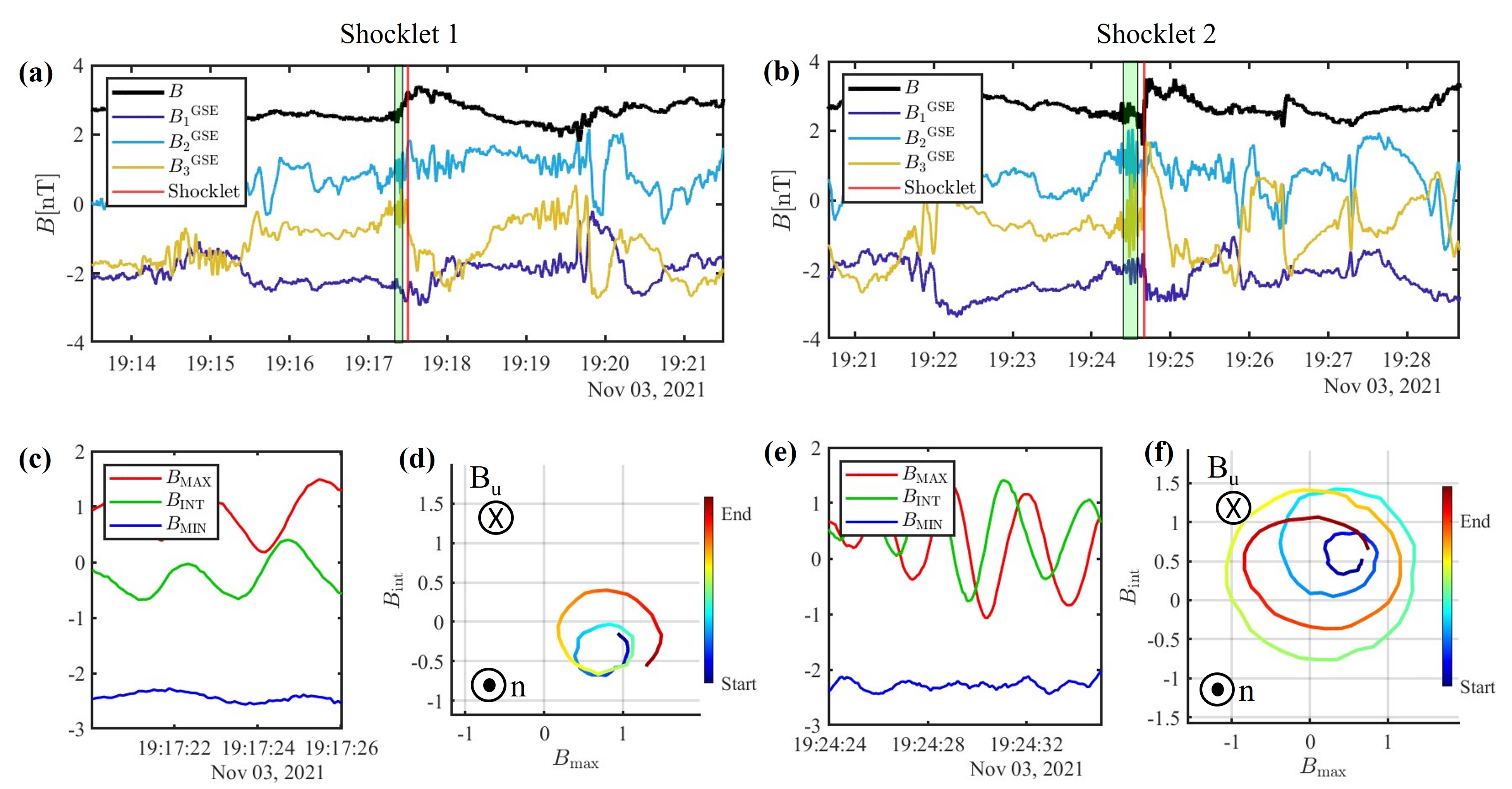}
    \caption{Panels (a)-(b): detailed view of the magnetic field around the upstream shocklets, identified by the vertical red line. The green shaded area indicates the whistler precursor. Panels (c)-(e): magnetic field components in the green shaded area, rotated to the minimum variance frame. Panels (d)-(f): magnetic field hodogram computed over the same time window, with the mean field and the MVA normal directions highlighted in the top and bottom left part of the plot, respectively. }
    \label{fig:fig4_hodo}
\end{figure*}

The IP shock had crossed Solar Orbiter at 14:04 of the same day. The overview plot on the left-hand side of Figure~\ref{fig:fig2_SoloVsWind} shows several interesting features. First of all, the magnetic field upstream of the shock reveals structuring over a broad range of frequencies, as we discuss in detail below. Particularly interesting is the discontinuity observed at around 12:30, associated with a slight increase of ion temperature upstream of the shock, and possibly a pre-conditioning of the incoming particle population. Another interesting feature of this IP shock crossing is related to the ion density increase observed by Solar Orbiter, that is not sharply rising at the time corresponding to the shock arrival, but instead grows smoothly deeper downstream, yielding a small value for the local (i.e., using averaging windows of order of a minute duration) evaluation of the gas compression ratio $\langle \mathrm{r} \rangle \sim 1.47$, while the magnetic compression ratio is larger (($\langle \mathrm{r_B} \rangle \sim 2.62$). Compression ratios show significant variations at each spacecraft, consistent with the high level of fluctuations observed for this shock. Finally, we note that the ion temperature increase observed at Solar Orbiter is rather small, with the temperature starting to rise a few minutes before the shock passage, probably due to shock-produced reflected particles injected upstream. Further properties of the shock, including the flux rope identified (not shown here) in the immediate shock downstream (and seen deeper in the shock downstream at Wind; not shown), are also interesting but beyond the scope of this paper.

\begin{figure*}
	\includegraphics[width=\textwidth]{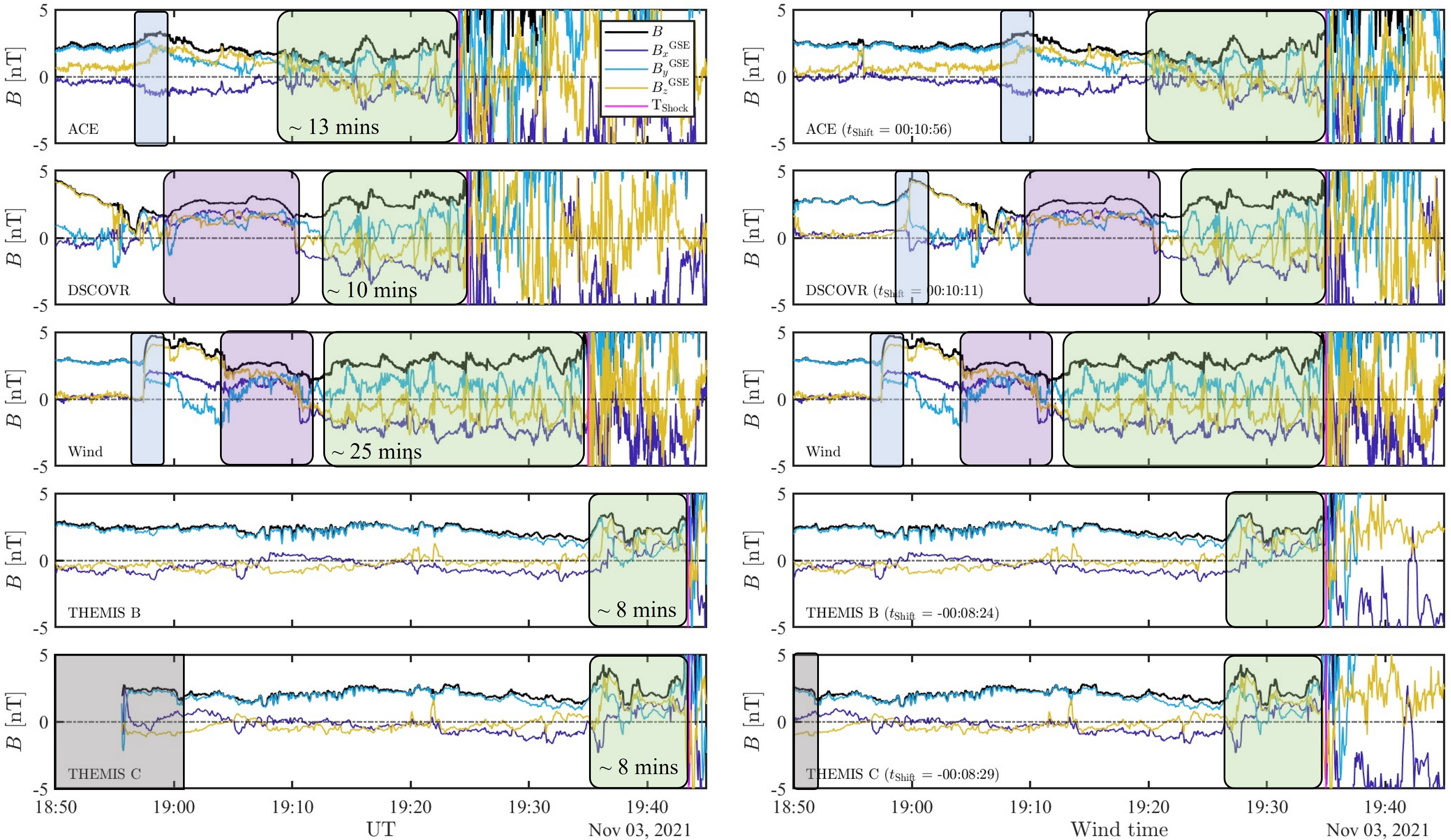}
    \caption{\emph{Left}: Magnetic field overview of the shock arrival at several satellites around L1. From top to bottom, are shown ACE, DSCOVR, Wind, THB and THC. The magenta line marks the shock crossing time for each spacecraft. The green shaded boxes mark the shocklets filled shock upstream, and the duration of such intervals is specified on each box. The grey shaded box represents a portion of THEMIS C data stream that may be affected by the lunar wake. The orange and blue boxes indicate structures seen at multiple spacecraft discussed below. \emph{Right}: Same as the left panels, but where the observation time has been shifted such that the shock crossing time at each spacecraft corresponds to the time at which the IP shock crosses the Wind spacecraft. }
    \label{fig:fig5_merge}
\end{figure*}

\subsection{Shocklets at Wind}
\label{subsec:shocklets}

In this section, the details of upstream shocklets are addressed using the Wind data, that has the highest cadence in magnetic field and plasma available data among the spacecraft that observe upstream shocklets. The 30 minutes upstream of the IP shock observed by Wind are characterised by the presence of several shocklets. Such structures can be clearly observed in the top panel of Figure~\ref{fig:fig3_shocklets_oview}, and are often associated with high frequency wave trains upstream their leading edges, as shown in Figure~\ref{fig:fig4_hodo}. The most evident shocklets have been highlighted in the Figure with orange shaded panels.

In panel b of Figure~\ref{fig:fig3_shocklets_oview}, we show the proton bulk flow speed along the shock normal, in the shock frame $U_{\mathrm{n}}^{\mathrm{sh}} \equiv \mathbf{U} \cdot \hat{n} - v_{\mathrm{sh}}$. We observe that the shocklets are associated with an effective deceleration of the upstream plasma, as it can be seen from the spikes in $U_{\mathrm{n}}^{\mathrm{sh}}$ (the clearest of them is highlighted by the black arrows in Figure~\ref{fig:fig3_shocklets_oview}). The noise level is higher for the proton density and temperature signals (panels c-d), however it is possible to see that plasma compression and heating associated with shocklets is resolved for some of them. 


Upstream shocklets are further investigated in Figure~\ref{fig:fig4_hodo}. Here, we show two examples of such structures in the shock upstream and highlight their details. These shocklets are highlighted by the black arrows in the overview Figure~\ref{fig:fig3_shocklets_oview}. Panels (a) and (b) of the Figure show zooms over $\sim$ 8 minute intervals of magnetic field magnitude and components (solid lines) around two shocklets. The vertical red line marks the upper half of the leading edge of the shocklet. We can see important differences between the Shocklet 1 and Shocklet 2. First of all, Shocklet 2 has a much sharper leading edge (with a rise time, i.e., the time between the backgorund upstream magnetic field value and its peak, of about 4 seconds), and shows strong structuring, including a well-developed high frequency wave train in the shocklet upstream and an overshoot/unsdershoot feature (immediately after the red line in Figure~\ref{fig:fig4_hodo} b) in the relaxation phase. On the other hand, Shocklet 1 is characterised by a  a larger rise time ($\sim$ 10 seconds) and less prominent structuring, with shorter high frequency wave packet upstream and a smoother relaxation phase ($\sim$ 90 seconds). We infer that, since Shocklet 1 is upstream of Shocklet 2, the former is still in the early phase of the process of steepening, and Shocklet 2 has evolved further. Such a consideration about the different stages in the structures' evolution, considering the fact that they are observed to be extremely close to each other ($\sim$ 10 minutes, corresponding to about 30 proton cyclotron times $T_{cp} \equiv 1/f_{cp}$, where $f_{cp}$ is the proton cyclotron frequency computed using the mean value for the upstream magnetic field of 3 nT) highlights their transient nature. 

The high frequency wave precursors have been highlighted with the green shaded areas in panels (a) and (b) of Figure~\ref{fig:fig4_hodo}. To further investigate the nature of these precursors, we performed a Minimum Variance Analysis \citep[MVA,][]{Paschmann2000} over the highlighted intervals. In these intervals, the intermediate to minimum eigenvalue ratio $\lambda_2/\lambda_3$ for the MVA matrix is large, with $\lambda_2/\lambda_3 \sim 15, \, 65$ for Shocklet 1 and 2, respectively. A larger value of these ratios can be achieved filtering the data \citep{Wilson2017}. The magnetic field components are projected to the minimum variance frame and shown in panels (c)-(e) of Figure~\ref{fig:fig4_hodo}. Here, it can be seen that the waves in the precursor have periods of about 2 seconds. Finally, the panels (d)-(f) of the Figure show hodograms for the intermediate and maximum variance magnetic field, showing that they are circularly right-hand polarised in the spacecraft frame. This is expected for whistler wave modes, and it is indicative of the dispersive nature of shocklets, consistent with other studies \citep[e.g.,][]{Hoppe1981,Wilson2009}. Note that these modes are often seen as left-handed in the terrestrial foreshock since they are trying to propagate against the solar wind in the spacecraft/shock frame.  In contrast, they are propagating with the solar wind here, thus they retain their intrinsic polarization.

\subsection{Multi-spacecraft observations of shocklets}
\label{subsec:multi-sc}

After a detailed characterisation of the shocklets observed by the Wind spacecraft, we address the spatial and temporal behaviour of them using the other nearby spacecraft. Figure~\ref{fig:fig5_merge} (left) displays the magnetic field measurements of 5 spacecraft: ACE, DSCOVR, Wind, THEMIS-B and THEMIS-C, respectively. All the spacecraft observe the IP shock passage (the times of crossing are reported in Table~\ref{tab:tab_event}). The shocklet field, i.e., the portion of shock upstream filled with shocklets, is highlighted by the green shaded boxes. Here, it is possible to see that shocklet fields have been simultaneously observed at ACE, DSCOVR and Wind, where the first two are upstream of Wind. It should be noted that, during the day analysed here, THEMIS B an C crossed the lunar wake (at times around 6 AM and 6 PM, respectively). The grey shaded panel in Figure~\ref{fig:fig5_merge} shows a portion of data from THEMIS C that may still be affected by the lunar wake.

To better understand the features of the upstream magnetic field for each L1 spacecraft, the measurements have been shifted in Figure~\ref{fig:fig5_merge} (right) such that the shock arrival is time is the same for each spacecraft (i.e., for each time series, the transformation $t_{\mathrm{Wind}} = t + t_{\mathrm{Shift}}$, where $ t_{\mathrm{Shift}}$ is the difference between the shock arrival time at each spacecraft and the shock arrival time at Wind). The vertical magenta line marks the shock transition time. The green shaded areas, show the shock upstream portion in which shocklets are observed. The shading scheme in the right-hand column is the same as the left-hand column of Figure~\ref{fig:fig5_merge}. The extent of these shocklet-filled portions of the shock upstream varies from $\sim$ 8 minutes to $\sim$ 25 minutes. As it is evident from  Figure~\ref{fig:fig5_merge}, shocklets appear to be persistent in the upstream upon its arrival at L1, but the finer details of their observations are extremely variable, possibly due to the rather small spatial and temporal scales (with respect to the inter spacecraft separations) characterising the shocklets.

\begin{figure}
	\includegraphics[width=\columnwidth]{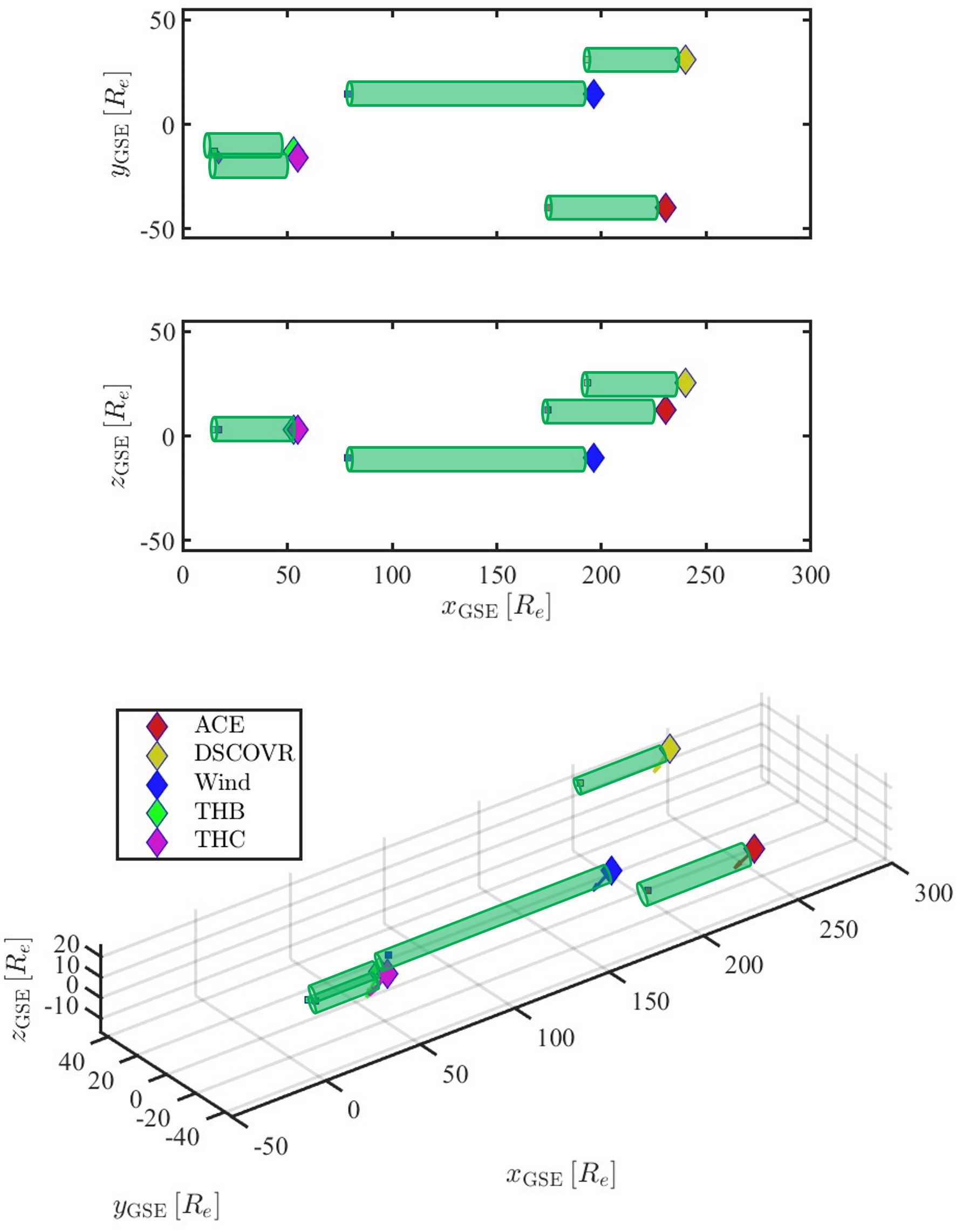}
    \caption{Top two panels: Spacecraft positions in the GSE coordinate systems (diamonds) and portion of their upstream filled with shocklets (green arrows). Bottom: Three-dimensional overview of spacecraft positions (diamonds). The colored arrows represent magnetic field measurements at Wind (blue) and THB (green), centered at the time of shock arrival at THB.}
    \label{fig:fig6_orbit_fill}
\end{figure}

It is worth noting another feature visible in the shock upstream in Figure~\ref{fig:fig5_merge}, namely the magnetic field structure seen around 18:55 at Wind, present also in the DSCOVR and ACE timeseries (see the blue shaded box in Figure~\ref{fig:fig5_merge}). The structure is characterised by a rise in field magnitude and a sharp change in the $y$ - $z$ components of the magnetic field. We note that no high frequency precursor is observed here. As we can see in the Wind temperature profile in Figure~\ref{fig:fig3_shocklets_oview}, this structure is also associated with a pre-heating of the incoming plasma. We speculate that the structure is probably a pre-existent solar wind feature ahead of the IP shock, due to  the presence of another structure upstream of the shocklet field, visible between 19:05 and 19:12 Wind time at Wind and 19:10 to 19:25 Wind time at DSCOVR and highlighted by the purple panels in Figure~\ref{fig:fig5_merge}. Furthermore, the average field direction downstream of the structure suggests that it is not connected to the portion of the shock front observed by DSCOVR and Wind. A very interesting behaviour noted for such a structure is that it has a shorter duration at Wind, compared to DSCOVR, possibly an effect due to the fact that Wind is below the ecliptic plane (see $x-z$ plane in Figure~\ref{fig:fig6_orbit_fill}). Another explanation for this shorter duration is that it is due to a compression for which the shocklet field, likely populated with energetic particles, is responsible. Another explanation for this shortening could be related to the three-dimensional nature of the structure and the direction at which Wind and DSCOVR are crossing it. Such speculation need corroboration coming from particle data, out of scope for the present work.

These simultaneous observations of upstream shocklets provide invaluable insights about their behaviour. First of all, combining the time series observations with the spacecraft locations, it is possible to infer a size for the portion of space filled with shocklets, thus relating the information obtained from the time series to a spatial information. Assuming a solar wind speed of 500 km/s (motivated by the Wind observations reported in Figure~\ref{fig:fig2_SoloVsWind}), the L1 observations reported in Figure~\ref{fig:fig5_merge} have been used to assess the spatial portion of the IP shock upstream where shocklets are present.

Such an information is shown in the top two panels of Figure~\ref{fig:fig6_orbit_fill}. Here, the spacecraft positions at the time of the shock arrival at Wind are displayed in the $x$ - $y$ and $x$-$z$ planes of the GSE coordinate system, together with the spatial region in which shocklets are observed (green shaded areas). The configuration is such that ACE and DSCOVR are the two most upstream satellites, THEMIS B and C are the two most downstream spacecraft. In Figure~\ref{fig:fig6_orbit_fill} it is possible to observe the larger spread in the $y$-direction than the $z$-direction for the spacecraft group. Finally, the bottom panel of the Figure shows a three dimensional overview of the spacecraft fleet in the GSE coordinate system, with the green shaded cylinders highlighting the shocklet field.

Neglecting shock curvature effects and considering the times in which DSCOVR and ACE simultaneously observe upstream shocklets, a lower limit for the extent of the shock front interacting with the upstream shocklets is about 71 $R_E$ (corresponding to $\sim$ 2200 proton skin depths $d_p$ using the average proton density upstream of Wind) in the $x$-$y$ plane of the GSE coordinate system and 40 $R_E$ ($\sim$ 406 $d_p$) in the $x$-$z$ plane, larger than the portion of the Earth's bow shock interacting with shocklets, being at most 30 $R_E$ \citep[e.g.,][]{VonAlftan2014}. With the same approach, we estimate a length for the upstream portion of space filled with shocklets using Wind observations. Again, assuming a solar wind speed of 500 km/s, the length of the upstream region filled the shocklets is at least 110 $R_E$ ($\sim$ 3500 $d_p$) long for the shock observed by Wind. Furthermore, the above results were projected in the shock normal frame using the shock normal vector computed at Wind, revealing that the shocklet field extends about 106 $R_E$ along the shock normal and 25 $R_E$ along the other two transverse directions.

\section{Conclusions}
\label{sec:discussion}

In this work, we studied the interesting behaviour of a CME-driven IP shock  observed near the Earth by ACE, DSCOVR, Wind, THEMIS B and THEMIS C (in order of shock arrival time). We focused on the shock crossing at Wind, where \emph{in-situ} analyses show that the shock was quasi-parallel and characterised by high fast magnetosonic and Alfvenic Mach numbers, unusual for IP shocks. These parameters are consistent with the ones observed for the shock crossings at the other spacecraft (see Table~\ref{tab:tab_event}). Upstream of this strong shock, we report very rare observations of shocklets, i.e., steep enhancements of magnetic field magnitude, with $\delta B/B_0 \lesssim 2$ and a typical time asymmetry between the rise and relaxation of the magnetic field signal. Using the Wind 3DP instrument, the presence of shocklets has been linked to an effective deceleration of the upstream plasma in the shock normal direction, thus highlighting their important role in pre-conditioning the incoming plasma for the shock transition. Performing a closer analysis of the shocklets, precursors of whistler waves have been identified in their upstream, a feature of their dispersive/transient nature. We note that the whistler precursor alone has been shown to affect the incident plasma \citep[][]{Chen2018,Hull2020,Wilson2009,Wilson2017}, and it is probably less influential than the entire shocklet structure, supporting the role of shocklets in the effective pre-conditioning of the upstream plasma.

Earlier in the day, at 14:04, the same shock was observed by Solar Orbiter at 0.81 AU from the Sun. Solar Orbiter was extremely well-aligned to Earth, making it an excellent proxy to investigate the time history of this IP shock. At Solar Orbiter, the shock is estimated to have an oblique geometry (\tbn $\sim 45^\circ$), and the shock parameters estimation is consistent with what observed at L1, namely high Mach number and shock speed. No shocklets were identified in the shock upstream at Solar Orbiter, probably due to the higher obliquity of the shock, together with the fact that the shock crossing happens in a very structured portion of the solar wind, due to the presence of magnetic structures both upstream and downstream, associated with changes in the plasma parameters. Another important ingredient is the shock time evolution: it is possible that the shock at Solar orbiter did not yet produce enough particles for efficient eave steepening upstream. We speculate that the higher shock obliquity, together with the structured nature of the shock upstream, create unfavorable conditions for upstream waves to steepen and grow into shocklet structures. 

However, observations of shocklets at IP shocks are extremely rare, with only two other cases reported in previous literature~\citep{Lucek1997,Wilson2009}, probably due to the fact that usually IP shocks are not as strong as the one observed here, and shock strength is known to play an important role in the generation of different upstream ion populations \citep[e.g][]{Savoini2015}. For these reasons, our focus is on the shocklets observations at L1 and near-Earth where, for the first time, we use a multi-spacecraft approach to study these interesting structures.

Upstream shocklets have been found at each of the five spacecraft mentioned above. The upstream field filled with shocklets at these spacecraft highlights again the transient nature of these structures, with variable duration at each spacecraft, even in presence of observations so close in time (see Figure~\ref{fig:fig5_merge}). The local shock obliquity estimates for this strong IP shock have a large range (9$^\circ \lesssim$ \tbn $\lesssim$ 64$^\circ$), while at smaller planetary bow shocks shocklets are associated with quasi-parallel geometries. This spatial/temporal variation at a large scale IP shock can offer an explanation for the previous surprising observation of shocklets at a similarly strong IP shock with local \tbn $\sim 68^\circ$ by \citet{Wilson2009}. This spacecraft configuration has also been used to address the portion of space filled with shocklets upstream of the IP shock, an important ingredient to consider when addressing several aspects of particle energisation and energy conversion at strong shocks. Using ACE and DSCOVR observations, a lower limit for such portion of space has been estimated to be of about 71 $R_e$  in the $x$-$y$ plane and 15 $R_e$ in the $x$-$z$ plane for the transverse directions of the GSE coordinate system, and of at least 110 $R_E$ along the $x$ GSE direction. These value correspond to about 2200, 406 and 3500 $d_p$, using the mean upstream proton density measured by the Wind spacecraft.

In a follow up work, the relation between the presence of shocklets and the production of energetic particles away from the shock front, as well as a characterisation of such structures from the point of view of scattering of suprathermal particles will be addressed, looking at the link between shock reflected particle distributions and mechanisms for wave steepening and shocklet formation. Under this point of view, it would be very interesting to study other strong IP shock events, that may become more common as we approach the solar maximum, using the capabilities of modern spacecraft such as Solar Orbiter, able to yield high resolution measurements of energetic particles.


\section*{Acknowledgements}

This work has received funding from the European Unions Horizon 2020 research and innovation programme under grant agreement No. 101004159 (SERPENTINE, www.serpentine-h2020.eu). D.T. is grateful to the SERPENTINE consortium for supporting this work. The work of H.H. is supported by the Royal Society award URF\textbackslash R1\textbackslash 180671. H.H. thanks discussions in the ISSI International Team 465 ``Foreshocks across the Heliosphere''. N.D. is grateful for support by the Turku Collegium for Science, Medicine and Technology of the University of Turku, Finland. 

\section*{Data Availability}

The Solar Orbiter data used in this study have been downloaded at soar.esac.esa.int. Data from all the other missions, used in this work, can be downloaded at  https://cdaweb.gsfc.nasa.gov/.



\bibliographystyle{mnras}
\bibliography{biblio} 






it can be placed in an Appendix which appears after the list of references.


\bsp	
\label{lastpage}
\end{document}